\documentclass[reqno,11pt]{amsart}
\usepackage[dvips]{changebar}
\usepackage[dvips]{color}
\usepackage{amssymb,amsthm}

\usepackage{a4}
\numberwithin{equation}{section}

\usepackage[mathscr]{eucal}

\usepackage{version}
\excludeversion{LONG}

\newcounter{mnotecount}[section]
\renewcommand{\themnotecount}{\thesection.\arabic{mnotecount}}

\newcounter{mymnotecount}[section]
\renewcommand{\themymnotecount}{\thesection.\arabic{mymnotecount}}
\newcommand{\mymnote}[1]
{\protect{\stepcounter{mymnotecount}}$^{\mbox{\footnotesize $%\!\!\!\!\!\!\,
\bullet$\themnotecount}}$ \marginpar{%\color{red}%
\raggedright\tiny\em $\!\!\!\!\!\!\,\bullet$\themymnotecount: #1} }
\renewcommand{\mymnote}[1]{}

\theoremstyle{plain}

\title{Helical solutions in scalar gravity}

\author[R. Beig]{Robert Beig${}^{\ddagger}$
} \email{robert.beig@univie.ac.at}
\address{Gravitational Physics, Faculty of Physics, University of
Vienna,
\newline
Boltzmanngasse 5, A-1090 Vienna, Austria}
\thanks{${}^{\ddagger}$ Supported in part by Fonds zur F\"orderung der
Wissenschaftlichen Forschung project no. P20414-N16.}
\author[B. Schmidt]{Bernd G. Schmidt} \email{bernd@aei.mpg.de}
\address{Max-Planck-Institut f\"ur Gravitationsphysik,
Albert-Einstein-Institut,
\newline
Am M\"uhlenberg 1, D-14476 Golm, Germany}

\begin{document}
\date{\today \ {\em File:\jobname{.tex}\,}}

\begin{abstract}
We construct solutions, for small values of $G$ and angular
frequency $\Omega$, of special relativistic scalar gravity coupled
to ideally elastic matter which have helical but no stationary or
axial symmetry. They correspond to a body without any symmetries in
steady rotation around one of its axes of inertia, or two bodies
moving on a circle around their center of gravity. Our construction
is rigorous, but modulo an unproved conjecture on the
differentiability of a certain functional.\\

\end{abstract}

\maketitle
\numberwithin{equation}{section}
\renewcommand{\theequation}{\thesection.\arabic{equation}}
\section{{\bf Introduction}}
 In Newton's theory of gravity there are solutions
describing two bodies which move at constant angular velocity on a
circle around their common center of gravity. The existence of such
solutions for fluid bodies was demonstrated by Lichtenstein  1922
\cite{li}. In Einstein's theory we do not expect such solutions to
exist, because such a system will emit gravitational radiation, and
so the two bodies will spiral towards each other.

However in 1992 Detweiler  and Blackburn \cite{de} conjectured that
such solutions should exist also in general relativity, provided
there is just the right amount of incoming gravitational radiation
to keep the bodies on their circle. J. Friedman, R.Price and
coworkers (\cite{fr}, \cite{pr} and references therein), followed
this idea and analysed various model systems mathematically and
numerically. There are many  conceptual and technical problems
involved. From the Schild solution \cite{sch} of electromagnetism,
which describes two point charges moving on a circle, it becomes
clear that such a system has infinite energy because the required
amount of incoming (and outgoing) radiation is unbounded. So, in GR
we expect that the ADM mass will be infinite. Consequently the
system will not by asymptotically flat in the sense of any of the
known definitions, which in turn makes it difficult to even define
such a system. We want the spacetime to admit a Killing vector which
behaves like $\partial_t+\Omega\partial_\phi$. The meaning of this,
however, is unclear if there is no asymptotic symmetry group. There
are ways out, Friedman proposes a definition in \cite{fr}.\\
 There is another subtlety: in  Minkowski space
the helical Killing vector $\partial_t+\Omega\partial_\phi$ is
timelike near the center and spacelike near infinity. Hence, the
metric on the quotient of the Killing vector changes signature. It
is completely unclear, how to deal with the Einstein equations in
such a situation.

All numerical model systems treated so far do not treat the 'bodies'
dynamically. In this paper we consider a special relativistic theory
in which the bodies are composed of some elastic material and
gravity is modeled by a scalar field. The main point of the paper is
to demonstrate  that solutions do exist in which the incoming
radiation keeps the bodies on their circle. In special relativity
there is no problem to define such systems. We just want the
solution to be invariant under the Killing vector
$\partial_t+\Omega\partial_\phi$. The scalar gravitational field
satisfies the wave equation in Minkowski space.

There is a case which is technically simpler than the 2--body
system. This is a rigidly rotating 'tri-axial body'. By this we mean
that the body does not have to have any symmetries. Again,
in Newtons theory such solutions exist \cite{bs1}. In GR, however,
the body should spin down because it radiates gravitational waves if
it is not axisymmetric with respect to the axis of the rotation.

The plan of the paper is as follows. In Section 2 we describe the
theory we are using: special relativistic elasticity formulated as a
Lagrangian field theory \cite{bs2} minimally coupled to a version of
scalar gravity. Section 3 discusses helical symmetry and derives the
equations for which we want to show existence. In Section 4 we
consider a triaxial body in helical motion and describe 'its
gravitational field' according to scalar gravity. We try to find a
solution with the right amount of incoming radiation by taking as
gravitational field (1/2 of) the sum of the retarded and advanced
solution of a given source motion. With any other combination our
method would break down. We want to show existence using the
implicit function theorem. Because the linearized operator has a
nontrivial range we first solve a 'projected system' first and only
then the full system.

In Section 5 we discuss in outline the helical  2--body problem for
which existence can be established by essentially the same method.
We were not able to show that no solution exists if we take just the
retarded solution of the wave equation.

There is a gap in our existence proof, because we do not demonstrate
that the 'gravitational force' depends differentiably on the
configuration, in the function spaces we are using for elasticity.
We consider this as a technicality. If this point can be cleaned up,
we are confident that even for complicated non linear systems such
as GR the idea that incoming radiation can balance outgoing
radiation works. Finally, we use elastic bodies only because in this
case we understand the boundary value problem. A corresponding
result should certainly also be true for fluids.

\section{{\bf Scalar gravity coupled to elasticity}}
Consider a Lagrangian field theory on Minkowski space $M$ space with
metric $\eta_{\mu\nu}=\mathrm{diag}(-1,1,1,1)$ with dynamical
variables given by a scalar field $V(x^\mu)$ descibing gravity and
and the fields $f^A(x^\mu)$ describing elasticity in Special
Relativity \cite{bs2}, which are viewed as maps $M \rightarrow
\mathcal{B}$ with $\mathcal{B}$, the so-called material space or
body, being a bounded domain in $\mathbb{R}^3$ with smooth boundary
$\partial \mathcal{B}$. The action is
\begin{equation}\label{1}
S = \frac{1}{2} \int \eta^{\mu \nu} V_{,\mu} V_{,\nu} \sqrt{-\eta}\; d^4x + 4\pi G
\int
\rho F(V)
\sqrt{-\eta}\; d^4x.
\end{equation}
We assume that $\mathcal{B}$ is endowed with an Euclidean metric
$\delta_{AB}$, where $A,B=1,2,3$ and that the scalar $\rho$ defining
the elastic material depends only on the principal invariants of the
matrix $H^A{}_B = H^{AC} \delta_{BC}$, where $H^{AB} = f^A{}_{,\mu}
f^B{}_{,\nu} \eta^{\mu\nu}$. This means, in standard language, that
we consider isotropic materials.
%\begin{equation}{}
%\rho=\rho(\eta^{\mu\nu}f^A_{,\mu}f^B_{,\nu})
%\end{equation}
%
The energy momentum tensor of the system is:

\begin{equation}\label{2}
T_{\mu\nu}^{\mathrm{tot}}= V_{,\mu} V_{,\nu} - \frac{1}{2} \eta_{\mu \nu} \eta^{\sigma
\lambda}
V_{,\sigma} V_{,\lambda} + 4\pi G  T_{\mu \nu} F,
\end{equation}
with $T_{\mu \nu}$ the elasticity energy momentum tensor from
\cite{bs2}.
\begin{equation}\label{3}
T_{\mu \nu}= \rho\, u_\mu u_\nu - \sigma_{\mu \nu}\,,
\end{equation}
where $u^\mu$ is the future-pointing vector field given by
\begin{equation}\label{4}
f^A{}_{,\mu} u^\mu = 0\,,\hspace{1cm} \eta_{\mu\nu} u^\mu u^\nu = -1
\end{equation}
The existence and uniqueness of $u^\mu$ is a regularity condition on
the fields $f^A$ we are considering. The quantity $\sigma_{\mu\nu}$
is given by

\begin{equation}\label{5}
- \sigma_{\mu\nu} = n \,\frac{\partial \epsilon}{\partial
H^{AB}} f^A{}_{,\mu} f^B{}_{,\nu}\,,
\end{equation}
where
\begin{equation}\label{11}
\rho=n \epsilon \ \ ,\
%H^{AB}=f^A_{,\mu}f^B_{,\nu}\eta^{\mu\nu}
\end{equation}
and $n$ is defined by
\begin{equation}\label{n}
\varepsilon_{ABC} f^A{}_{,\mu}(x) f^B{}_{,\nu}(x)
f^C{}_{,\lambda}(x) = n(x) \ \epsilon_{\mu\nu\lambda\sigma} u^\sigma
(x)
\end{equation}
The equation for $V$ is
\begin{equation}\label{6}
\Box V - 4\pi G \rho F'=0.
\end{equation}
To obtain a linear equation we choose
\begin{equation}\label{7}
F(V)=1+V
\end{equation}
We have
\begin{equation}\label{8}
\nabla^\nu\, T_{\mu \nu}^{\mathrm{tot}}=0.
\end{equation}
The equations (\ref{2},\ref{8}) imply
\begin{equation}\label{9}
0=\rho  V_{,\mu}+T_\mu{}^\nu V_{,\nu}+(1+V)(\rho u_\mu u^\nu-\sigma_\mu{}^\nu{})_{,\nu}
\end{equation}
Or ($h_\mu{}^\nu =\delta_\mu{}^\nu + u_\mu u^\nu$)
%
%\begin{equation}\label{10}
%0=(\rho  h_\mu{}^\nu + \sigma_\mu{}^\nu )V_{,\nu}+(1+V)(\rho u_\mu
%u^\nu+\sigma_\mu{}^\nu{})_{,\nu}
%\end{equation}
%
%\begin{equation}\label{11}
%\rho=n \epsilon \ \ ,\
%H^{AB}=f^A_{,\mu}f^B_{,\nu}\eta^{\mu\nu}
%\end{equation}
%
%which implies
%
%\begin{equation}\label{12}
%\sigma_{\mu\nu}=n \tau_{AB}f^A{}_{,\mu}f^B{}_{,\nu}\ ,\hspace{0.5cm}\
%\tau_{AB}=2\frac{\partial \epsilon}{\partial H^{AB}}
%\end{equation}
%
%Finally we obtain
%
\begin{equation}\label{13}
0=\left[\rho\, h_\mu{}^\nu - \sigma_\mu{}^\nu
\right]V_{,\nu} + \left(1+V \right)(\rho u_\mu
u^\nu-\sigma_\mu{}^\nu{})_{,\nu}
\end{equation}
Note that in the calculation of $\sigma_\mu{}^\nu{}_{,\nu}$ in terms
of the $f^A$  the first term in $\rho$ does not contribute.
Furthermore, both the 'force term' (involving $V_\nu$) in (\ref{13})
and the remaining term is orthogonal to $u^\nu$ (for the latter this
is e.g. shown in \cite{bs2}).

We have a PDE system for $V,f^A$. This has a Newtonian limit which is treated in Sect.5 of
\cite{bs1}.
\section{{\bf Helical motion}}
\setcounter{equation}{0}
We assume that the material flow is
parallel to the helical Killing vector $\partial_t+\Omega
\partial_\phi$, i.e. that
\begin{equation}\label{13'}
f^A{}_{,\mu} (\partial_t+\Omega
\partial_\phi)^\mu = 0
\end{equation}
 and that $V$ is invariant under $\partial_t+\Omega
\partial_\phi$, i.e.
\begin{equation}\label{13''}
V_\mu (\partial_t+\Omega
\partial_\phi)^\mu = 0
\end{equation}
For concreteness we choose coordinates $(\tau,y)$, so that
$\tau = 0$ coincides with the hyperplane $t=0$ with $y$ Euclidean coordinates thereon
and $\partial_t + \Omega \partial_\phi = \partial_\tau$. Explicitly the transformation is given by
\begin{equation}\label{13'''}
x^1 = y^1 \cos \Omega \tau - y^2 \sin \Omega \tau,\,x^2 = y^1 \sin \Omega \tau + y^2 \cos \Omega \tau,\, x^3 = y^3,\, t=\tau
\end{equation}
In these coordinates the configuration $f^A$ can be written as functions $\hat{f}^A$ of $y$, and similarly
for $V$. By slight abuse of notation we will omit the bar in what follows. Now Eq.(\ref{13'}) implies that the quantities $n, \rho$ and $\sigma_\mu{}^\nu$ are all
also invariant under $\partial_t+\Omega
\partial_\phi$ and $\sigma_\mu{}^\nu (\partial_t+\Omega
\partial_\phi)^\mu = 0$. Using the orthogonality pointed out at the end of
the previous section and the expression for the projection of
$\sigma_\mu{}^\nu{}_{,\nu}$ on the quotient calculated in the paper
\cite{bs1}, we obtain
\begin{equation}\label{14}
0=\left[\rho h_j{}^i - \sigma_j{}^i \right]D_iV
+\left(1+V\right)\left[- e^{-U}D_i(e^{U}\sigma_j{}^i)+\rho D_jU \right] \ ,
\end{equation}
where $- e^{2 U}$ is the norm of the helical Killing vector
given by
\begin{equation}\label{14'}
e^{2 U} = 1 - \Omega^2 r^2\,,\hspace{0.7cm}r^2 = (x^1)^2 + (x^2)^2\,,
\hspace{0.7cm}r^2 \Omega^2 < 1
\end{equation}
$h_{ij}$ is the metric on the quotient of $\partial_t+\Omega
\partial_\phi$ and $D_i$ the covariant derivative with respect to $h_{ij}$. Note that, for $\Omega = 0$, $h_{ij}$ equals $\delta_{ij}$.
Note also that now $H^{AB} = f^A{}_{,i} f^B{}_{,j} h^{ij}$. We rewrite (\ref{14}) as
\begin{equation}\label{15}
D_i\left[\left(1+V\right) e^{U} \sigma_j{}^i\right]- \rho\, e^U
\left[D_j V + \left(1 + V\right) D_jU \right] = 0
\end{equation}
The second term and third term in (\ref{15}) is respectively the
gravitational and centrifugal force. We write (\ref{15}) as
\begin{equation}\label{16}
0= D_i\left[\left(1+V\right)e^{U}\sigma_j{}^i\right]+W_j+Z_j\,,
\end{equation}
and
\begin{equation}\label{20}
- W_j= e^U \rho \,D_j V
\end{equation}
with
\begin{equation}\label{21} - {Z}_j = e^U \rho
\left(1+V\right)D_j U \,,
\end{equation}
and
\begin{equation}\label{22}
D_j U= \Omega^2 \left(1 -\Omega^2 r^2 \right)^{-\frac{1}{2}}
\frac{1}{2} \,D_j\, r^2
\end{equation}
\section{Triaxial Rotating body}
The first case we want to consider is a triaxial rotating body
coupled to scalar gravity. We repeat that 'tri-axial' means that the
set $\mathcal{B}$ need not have any symmetries. Let us denote the
corresponding relaxed configuration by $\bar\Phi$. We take for the
gravitational field the symmetric (i.e. one half the sum of the
retarded and advanced) solution of the scalar wave equation. Note
that this makes sense globally in Minkowski space. But we will only
make use of that solution inside the support of the body, where $r^2
\Omega^2 < 1$. For a source $\rho$ invariant under helical motion we
have from the Appendix that
%
%\begin{equation}\label{17}
%\rho(t,\phi,\theta,r)=\hat\rho(\phi-\Omega t,\theta,r)
%\end{equation}
%
%and
%\begin{eqnarray}\label{18}
%V(t,\phi,\theta,r)&=&-{G\over
%2}\int{\hat\rho(\phi'-\Omega(t-c^{-2}|\vec x-\vec
%x|),\theta',x')\over|\vec x-\vec  x'|}\,d^3 x' - {}
%\nonumber\\
%&&{}\!\!-{G\over 2}\int{\hat\rho(\phi'-\Omega(t+c^{-2}|\vec x-\vec
%x'|),\theta', x')\over|\vec x-\vec x'|}\,d^3 x' {}
%\end{eqnarray}
%%
%At $t=0$ we obtain
%%
%\begin{eqnarray}\label{19}
%V(0,\phi,\theta,r)&=&-{G\over 2}\int{\hat\rho(\phi'+\Omega
%c^{-2}|\vec x -\vec x'|,\theta', x')\over|\vec x-\vec  x'|}\,d^3 x'
%- {}
%\nonumber\\
% &&{}\!\!-{G\over 2}\int{\hat\rho(\phi'-\Omega c^{-2}|\vec
%x-\vec x'|,\theta', x')\over|\vec x-\vec x'|}\,d^3 x '{}
%\end{eqnarray}
%
\begin{equation}\label{17}
V(y) = G \int H(y,y')\rho(y')  \sqrt{-\xi^2(y')\,h(y')}\,dy'\,,
\end{equation}
where $H(y,y')$ is given by (\ref{A}). Clearly this function is even in $\Omega$. Expanding $V$ in $\Omega$
only even powers occur and we have
\begin{equation}\label{18}
V=G\tilde V= V^N + \Omega^2 V^2 + \dots=G(\tilde V^N+\Omega^2\tilde
V^2+\dots)
\end{equation}
where $V^N$ is the Poisson integral of the source.
\subsection{ Solution of the projected system}
The fundamental equation (with $V$ still chosen half the sum of
retarded and advanced potential defined by $\rho$) is
\begin{equation}\label{19}
0=D_i\left[\left(1+V\right)e^U \sigma_j{}^i\right]+G\tilde
W_j+\Omega^2\tilde Z_j
\end{equation}
with
\begin{equation}\label{201}
W_j = G \tilde W_j
\end{equation}
and
\begin{equation}\label{211}
{Z}_j =\Omega^2\tilde W_j
\end{equation}
Equation (\ref{19}) forms a quasilinear second-order system of partial differential equations
for the functions $f^A$, which has to be solved subject to the (Neumann-type) boundary condition
\begin{equation}\label{23}
\sigma_i{}^j n_j|_{f^{-1}(\partial \mathcal{B})} = 0
\end{equation}
Due to the presence of a free boundary we will  also need the material form of (\ref{19}),
where the dependent variable $f(y)$ is replaced
by its inverse $\Phi(X)$ with $X \in \mathcal{B}$. Then, with $\bar W_j=n^{-1}\tilde W_j (\phi(X)))$
and $\bar Z_j=n^{-1}\tilde Z_j (\phi(X)))$, there results
\begin{equation}\label{24}
0=F_j(\Phi,\Omega^2,G)=\nabla_A\left[\left(1+G\bar
V\right)e^U \sigma^A{}_j\right]+G\bar
W_j+\Omega^2\bar Z_j
\end{equation}
where $\sigma^A{}_j$ is the 'first Piola stress' given by $\sigma^A{}_j = n^{-1} f^A{}_i \sigma^i{}_j$.
The boundary condition now takes the form
\begin{equation}\label{25}
\sigma^A{}_i n_A |_{\partial \mathcal{B}}=0
\end{equation}
We have two parameters in the problem, namely $\Omega^2$ and $G$.\\
Now one has to define an appropriate space for $\Phi(X)$, which will
be a neighbourhood of the identity map in a suitable Sobolev space,
we refer for details to the paper \cite{abs}. Let $\bar\Phi$ be a
relaxed, i.e. stressfree configuration, which we take to be the
identity map. We also take $\mathcal{B}$ to be contained in the set
$[(X^1)^2 + (X^2)^2] \Omega^2 < 1$, whence, for deformations
$\Phi(X)$ sufficiently close to the identity, the set
$\Phi(\mathcal{B})$ is inside
the light cylinder $[(y^1)^2 + (y^2)^2] \Omega^2 < 1$.\\
We next require that the function $\epsilon(H^{AB})$ in (\ref{11})
satisfies
\begin{equation}\label{26}
\epsilon(\delta^{AB}) = \mathring{\epsilon} = \mathrm{const} >
0\,\,,\hspace{1cm}\frac{\partial \epsilon}{\partial
H^{AB}}\,\Bigg{|}_{H^{AB} = \delta^{AB}} = 0
\end{equation}
Then we have a solution $F_j(\bar\Phi,0,0)=0$ of (\ref{24}). Assume that furthermore
\begin{equation}\label{27}
\left(\frac{\partial^2 \epsilon}{\partial H^{AB}\partial H^{CD}}\right)\Bigg{|}_{H^{EF} = \; \delta^{EF}} =
\lambda \,\delta_{AB} \delta_{CD} + 2 \mu\, \delta_{C(A} \delta_{B)D}
\end{equation}
with the constants $\mu,\lambda$ obeying $\mu > 0,\,3 \lambda + 2
\mu > 0$. The linearization of $F_j$ at $(\bar{\Phi},0,0)$ is the
standard operator of flat-space linear elasticity on $\mathcal{B}$,
which has to be considered together with the linearized form of the
boundary condition (\ref{25}). It is well-known (see e.g.
\cite{abs}) that this is an elliptic operator with finite
dimensional kernel and range. The latter is given by fields
$l_i(X)$, such that $\int_\mathcal{B} \eta^A(X)\, \delta^i{}_A\,
l_i(X)\, d^3X = 0$ for all Euclidean Killing vectors $\eta^A$ on
$\mathcal{B}$. One can like in \cite{abs} define a projection
$\mathbb{P}$ onto the range of the linearized operator and study the
projected system
\begin{equation}\label{28}
0=\mathbb{P} F_i(\Phi,\Omega^2,G)=\mathbb{P} \left\{
\nabla_A\left[\left(1+G\bar V\right)e^{\frac{U}{c^2}}\sigma^A{}_i\right]+G\bar
W_i+\Omega^2\bar Z_i \right\}
\end{equation}
Assuming differentiability of this nonlinear map  we can apply the
implicit function theorem. Differentiability is standard for the
$\bar Z_i$ - term (see \cite{bs1}). For the $\bar W_i$ - term, which
contains the retarded + advanced scalar field, we have to leave this
as a conjecture. The corresponding result for the Newtonian gravitational field has been proved in \cite{bs4}.\\
The kernel of the linearized operator consists of Killing vectors of flat Euclidean space.
We restrict the deformations $\Phi$ to ones of the form
\begin{equation}\label{29}
\Phi^i(X^A,G,\Omega^2,a^i,b^{jk})=\delta^i{}_A X^A+ a^i
+b^{ik}\delta_{kA}X^A + \tilde\Phi
\end{equation}
where $a^i$ and $b^{ij}$ are constants with $b^{ij} = b^{[ij]}$ and $\tilde\Phi$ is in a complement of the kernel which contains $\bar\Phi$.
Thus  we obtain a unique solution of the projected equations for each value of $a^i,b^{ik}$  defining an infinitesimal motion.
Hence we know  $\tilde\Phi(X^A,G,\Omega^2,a^i,b^{ik})$.

\subsection{{\bf Equilibration}}
To obtain a solution of the full equations we will construct a family
$(a^i(G,\Omega^2),b^{jk}(G,\Omega^2))$ such that
\begin{equation}\label{30}
\Phi^i(X^A,G,\Omega^2,(a^i(G,\Omega^2),b^{jk}(G,\Omega^2)))
\end{equation}
satisfies the equation and the boundary condition. We will determine this family by solving $(\mathbb{I} - \mathbb{P})F_i=0$.
This condition is equivalent to
\begin{equation}\label{31}
0=\tilde N_{(\alpha)}(G,\Omega^2,C)=\int_\mathcal{B}
\xi^i_{(\alpha)} \left\{
\nabla_A\left[\left(1+G\bar V\right)e^U \sigma^A{}_i\right]+G\bar
W_i+\Omega^2\bar Z_i\right\}d^3X
\end{equation}
 for all Killing vectors $\xi^i_{(\alpha)}$ with $\alpha = 1,...6$ of the flat metric on the body,
 $C = (a^i, b^{jk})$ and $\Phi^i(X^A,G,\Omega^2,C)$ is inserted in
the integrand. Note that $\xi^i_{(\alpha)} = a'^i + b'^i{}_j y^j$
has to be composed with $\Phi$, so the rotational terms depend on
$\Phi^i(X^A,G,\Omega^2,C)$. To have only one parameter we assume
$\omega^2=\kappa G$ with $\kappa$ a positive constant and, with this
in mind, define the "normalized force map" as
\begin{equation}\label{32}
N_{(\alpha)}(G,C)=G^{-1}\int_\mathcal{B} \xi^i_{(\alpha)}\left\{
\nabla_A\left[\left(1+G\bar V \right)e^U \sigma^A{}_i\right]+G\bar
W_i+\Omega^2\bar Z_i\right\}d^3X\
\end{equation}
and replace (\ref{31}) by
\begin{equation}\label{31'}
N_{(\alpha)}(G,C)=0
\end{equation}
This is well defined because both the forces and $\sigma^A_i$ have a
factor $G$.\\
In analyzing the 'equilibration condition' (\ref{31'}) we show,
first of all, that (\ref{31'}) is indeed satisfied for arbitrary
$G$, when $\xi_{(\alpha)}$ is $\partial_{y^3}$ or $y^2
\partial_{y^1} - y^1 \partial_{y^2}$ or a constant linear combination thereof. To prove this, it is simplest to look at the spatial
version of (\ref{31'}), namely (see (\ref{19}))
\begin{equation}\label{31''}
G^{-1} \int_{\Phi(\mathcal{B})} \xi^j_{(\alpha)}
\left\{D_i\left[\left(1+V \right)e^U \sigma_j{}^i\right]+G\tilde
W_j+\Omega^2\tilde Z_j \right\} \sqrt{h}\, d^3 y = 0
\end{equation}
Now, the first term in (\ref{31''}) is zero: use integration by
parts, the Killing equation for each of these two vectors and the
vanishing of the boundary term by virtue of (\ref{23}). For the
third contribution already the integrand is zero, again by the
symmetries of $U$. Finally, the second term is zero: this amounts to
the statement, proven in the Appendix, that, for a body which
rotates rigidly around
 the $y^3$ - axis, the $y^3$ - component of both the force and torque due to its own gravitational
 field is zero, provided this field is given by the symmetric solution of the wave equation (\ref{6}).\\
We thus only consider
\begin{equation}\label{31'''}
N_{(\alpha')}(G,C')=0
\end{equation}
where $C'=(a^1,a^2,a^3\!=\!0,b^{12}\!=\!0,b^{13},b^{23})$ and
$\xi_{(\alpha')}$ to be the collection
\begin{equation}\label{31iv}
\{\xi_{(1)},\xi_{(2)},\xi_{(13)},\xi_{(23)}\} = \{\partial_{x^1},
\partial_{y^2}, y^3 \partial_{y^1} - y^1 \partial_{y^3}, y^3
\partial_{y^2} - y^2 \partial_{y^3}\}\,.
\end{equation}
Suppose $N_{(\alpha')}(0,0)=0$ and $\frac{\partial N_{(\alpha')}}{\partial C'}(0,0)$ is invertible. Then, by the (finite-dimensional) implicit
function theorem, for sufficiently small $G$ there exists a function $C'(G)$ such
that $N_{(\alpha')}(G,C'(G))=0$, and we have a solution of our
problem.\\
Calculating $N(0,0)$ is the same as $\frac{\partial\tilde N}{\partial G}$. The first term in the integrand gives
$\partial_A\delta\sigma^A_i$ which is equilibrated as was discussed in Eq.(4.2) of \cite{abs}. So we have only to consider the  force terms.
\begin{equation}\label{33}
N_{(\alpha')}(G,C')=\int_\mathcal{B} \xi^i_{(\alpha')}\left(
 \bar W_i +\kappa\bar Z_i\right)d^3X\ .
\end{equation}
We obtain
\begin{equation}\label{34}
N_{(\alpha')}(0,C')=-\int_\mathcal{B} \left[\xi^1_{(\alpha')}(\Psi)
\Psi^1+\xi^2_{(\alpha')}(\Psi)\Psi^2\right]d^3 X + \int_\mathcal{B}
\xi^i_{(\alpha')} (\Psi) \bar{W}^N_i (\Psi)\, d^3X
\end{equation}
where $\Psi=\delta^i_A X^A+ a^i +b^{ik}\delta_{kA}X^A $ with $a^3 =
b^{12} = 0$. The $\bar W_i^N$ - term is simply the Newtonian
gravitational self-force (resp. self-torque), and so its contribution to
the integral vanishes (see e.g. example (i) in the Appendix). We are thus left with
\begin{equation}\label{35}
N_{(\alpha')}(0,C')=-\int_\mathcal{B} \left[\xi^1_{(\alpha')}(\Psi)
 \Psi^1+\xi^2_{(\alpha')}(\Psi)\Psi^2 \right]\,
d^3X\ ,
\end{equation}
where
\begin{equation}
N_{(1)}(0,C')=-\int_\mathcal{B} \left(X^1+a^1+b^{13}X^3\right) d^3X
\end{equation}
\begin{equation}
N_{(2)}(0,C')=-\int_\mathcal{B} \left(X^2+a^2+b^{23}X^3\right) d^3X
\end{equation}
\begin{equation}
N_{(13)}(0,C')=-\int_\mathcal{B} X^3\left(X^1+a^1+b^{13}X^3\right)
d^3X
\end{equation}
\begin{equation}
N_{(23)}(0,C')=-\int_\mathcal{B} X^3\left(X^2+a^2+b^{23}X^3\right)
d^3X
\end{equation}
From here we infer that $N_{(\alpha')}(0,0)$ is zero iff (i) the
rotation axis in the reference estate goes through the center of
mass and (ii) coincides with one of the axes of inertia. In fact, we
assume that the center of mass is at the origin, rather than just
(i). With these assumptions minus the matrix $\frac{\partial
N_{(\alpha')}} {\partial C'}\,(0,0)$ is given by
\begin{displaymath}
\left(\begin{array}{cccc} \mathcal{V} &  0 & 0 & 0\\
0 &   \mathcal{V}   &    0 & 0\\
0 & 0 &     \mathcal{W}& 0  \\
0 &   0 &  0 &  \mathcal{W}
\end{array} \right)
\end{displaymath}
%\begin{equation}
%\left(  \matrix{
%                 \mathcal{V} & 0 & 0 & 0  \cr
%                 0 & \mathcal{V} & 0 & 0 \cr
%                 0 & 0 & \mathcal{W} & 0  \cr
%                 0 & 0 & 0 & \mathcal{W}  \cr
%                }\right)
%\end{equation}
where
\begin{equation}
\mathcal{V}=\int_\mathcal{B} d^3X \,\hspace{1cm}
\mathcal{W}=\int_\mathcal{B} (X^3)^2d^3X
\end{equation}
So $\frac{\partial N_{(\alpha')}}{\partial C'}\,(0,0)$ is invertible,
and our argument is complete.
\section{Two bodies in circular motion}
In this section we consider two identical bodies in circular motion
around their common center. To simplify matters we assume that the
support of the matter in the relaxed configuration consists of two
spherical balls with centers at $(-L,0,0)$ and $(L,0,0)$ on the
quotient metric. We have as discrete symmetry the reflections
$(y^1,y^2,y^3)\to (y^1,-y^2,y^3)$ and $(y^1,y^2,y^3)\to
(y^1,y^2,-y^3)$. For an isotropic stored energy functions we know
that the solution will also have this symmetry. Therefore we
consider only configurations with the property
\begin{equation}
\Phi^1(X^1,X^2,X^3)=\Phi^1(X^1,-X^2,X^3)=\Phi^1(X^1,X^2,-X^3)
\end{equation}
$$
\Phi^1(X^1,X^2,X^3)=-\Phi^1(X^1,-X^2,X^3)=\Phi^1(X^1,X^2,-X^3)
$$
$$
\Phi^1(X^1,X^2,X^3)=\Phi^1(X^1,-X^2,X^3)=-\Phi^1(X^1,X^2,-X^3)
$$
One is now able to show the following:
\bigskip

1.) For such configurations the gravitational field "ad+ret"
inherits this symmetry. So does the centrifugal force.

\bigskip

2.)
For an  isotropic stored energy also the stress tensor and its divergence inherit this symmetry because it
is an isometry of the quotient metric.

\bigskip
3.) All these properties imply that all equilibration integrals,
with the exception of that for the $y^1$ -- translation  Killing
vector, are  satisfied for any configuration with these symmetries.

\bigskip
4.) If we assume further that $y^1\to -y^1 $, which exchanges the
two bodies, is a symmetry we have only to consider the elastic
equations for one body.

\bigskip
5.) As usual, we first solve the projected equations with a
parameter $a^1$, the translation in the $y^1$--direction, and have
then to study the bifurcation equation $N(G,a^1)=0$. It turns out
that $N(0,a^1)=0$ is the same as in the Newtonian problem which we
solved in \cite{bs3}. We find  a reference configuration with
$N(0,0)=0$ and $\frac{\partial N}{\partial a}(0,0) =0$ and have thus
solved the 2-- body problem.

\section{Appendix}
In this Appendix we prove a result which contains as a special case
the statement that the gravitational contribution to the
equilibration integrals (\ref{32}) vanishes, when the Killing vector
is taken to be $\partial_{x^3}$ or $\partial_\phi = - x^2
\partial_{x^1} + x^1 \partial_{x^2}$.\\
Let $\rho(t,x)$ be a source and
\begin{equation}
V(t,x) = \int G(t-t';x-x')\,\rho(t',x')\, dt'\, d^3x'
\end{equation}
be the field, where $G$ is 'some unique' Green function (in fact:
distribution) of $\Box \Phi = \rho$. By 'some unique Green function'
we mean a Green function sharing the symmetry of the background
(i.e. Minkowski space in our case), such
as the retarded or advanced Green function. We assume that $\rho$ has compact support in space.\\
Let us next suppose that the source $\rho$ is invariant under the flow of some timelike Killing vector $\xi^\mu\partial_\mu$. Using coordinates comoving with $\xi$, i.e. $(\tau;y)$ so that $\xi^\mu \partial_\mu = \partial_\tau$, Eq.(1) takes the form
\begin{equation}
V(\tau,y) = \int H(y,y')\rho(y')  \sqrt{-\xi^2(y')\,h(y')}\,d^3y'\,,
\end{equation}
where $H (y,y') = \int G(\tau - \tau';y,y') d\tau'$, $\xi^2 = g_{\mu\nu}\xi^\mu\xi^\nu$ and $h$ is the determinant of the metric on the quotient of $\xi^\mu$. (The square root in (2) is $\sqrt{- g}$ in the comoving system.) Clearly $V$ is actually independent of $\tau$.\\
Let now $\eta$ be a spacelike Killing field commuting with $\xi$,
i.e. projecting to $\eta = \eta^i (y) \partial_i$ on the quotient.
Consider the expression $F_\eta$ given by
\begin{equation}\label{plus}
F_\eta = \int \rho(y)\,[\eta^i \partial_i \,H(y,y')]\,\rho(y')\, \sqrt{-\xi^2(y)\,h(y)}
\,\sqrt{-\xi^2(y')\,h(y')}\;d^3y\;d^3y'
\end{equation}
This is up to sign the equilibration integral
$\int_{f^{-1}(\mathcal{B})} \eta^i W_i \,\sqrt{h}\, d^3y$. Since the
(unique) Green function shares the symmetry of the background there
has to hold (operate on both arguments with the symmetry and
linearize)
\begin{equation}\label{one}
[\eta^i (y) \partial_i + \eta^{i'} (y') \partial_{i'}]\, H(y,y') = 0
\end{equation}
(Note this relation holds irrespective of whether we take the retarded, advanced or any other combination.)
Furthermore $\eta^i \partial_i$ is a Killing vector in the quotient space, so that in particular
\begin{equation}\label{two}
\partial_i (\sqrt{h}\, \eta^i) = 0
\end{equation}
Also there holds
\begin{equation}\label{three}
\eta^i \partial_i\, \xi^2 = 0
\end{equation}
We use (\ref{one},\ref{two},\ref{three})  and integration by parts in $y'$ to obtain (the two
minus-signs arising in the process cancel)
\begin{equation}
F_\eta = \int \rho(y) \,H(y,y')\,[\eta^{i'}(y')\partial_{i'}\,\rho(y')]\sqrt{-\xi^2(y)\,h(y)}
\,\sqrt{-\xi^2(y')\,h(y')}\,d^3y\,d^3y'
\end{equation}
But we can also perform partial integration in the $y$-variable to find that
\begin{equation}\label{minus}
F_\eta = - \int [\eta^i (y)\partial_i \,\rho(y)]\, H(y,y')\, \rho(y') \sqrt{-\xi^2(y)\,h(y)}
\,\sqrt{-\xi^2(y')\,h(y')}\,d^3y\;d^3y'
\end{equation}
We now assume that the Green function is the symmetric one, so that
$G$ is invariant under the joint interchange of $(\tau,\tau')$ and
$(y,y')$. It follows that $H(y,y')= H(y',y)$. Whence, by
(\ref{plus},\ref{minus}), we find that  $F_\eta$ is zero.
We illustrate the previous result by two examples:\\
(i) static symmetry: Here $\xi$ is given by $\xi^\mu \partial_\mu = \partial_t$, and the, say retarded,
Green function in adapted coordinates takes the well-known form ($\tau = t$, $y = x$):
\begin{equation}
4 \pi G_{\mathrm{ret}}(\tau - \tau';y,y') = \frac{\delta(\tau - \tau' - |y - y'|)}{|y - y'|}
\end{equation}
Furthermore
\begin{equation}
4 \pi H(y,y') = \frac{1}{|y - y'|}
\end{equation}
i.e. $H(y,y')$ is the Poisson kernel. (This result is of course independent on which combination of the retarded and advanced Green function has been chosen.) The quotient metric $h$ is the
Euclidean one and spacetime Killing vectors projecting to the
quotient in this case are simply all Euclidean ones in $y$ - space.
We have thus recovered the well-known Newtonian result, that the
force and torque on a static body due to its own gravitational field are zero.\\
(ii) Helical symmetry (see \cite{bhs}): Here $\xi$ is given by $\xi^\mu \partial_\mu = \partial_t +
\Omega \partial_\phi$, and the retarded Green function takes the form ($\tau = t$, $\mu = \phi - \Omega t$, $z = x^3$)
\begin{align}
4 \pi G_{\mathrm{ret}}(\tau-\tau';y,y') &=
\nonumber\\
&\!\!\!\!\!\!\!\!\!\!\!\!\!\!\!\!\!\!\!\!\!\!\!\!\!\!\!\!\!\!\!\!\!\!\!\!\!\!\!\!\!\!\!\!\!
 =\frac{\delta\left(\tau - \tau' - \sqrt{r^2 + r'^2 - 2 r r' \cos[\mu - \mu'+ \Omega(\tau - \tau')] + (z-z')^2}\,\right)}
{\sqrt{r^2 + r'^2 - 2 r r' \cos[\mu - \mu'
+ \Omega(\tau - \tau')] + (z-z')^2}}
\end{align}
whence (assuming $r^2 \Omega^2 < 1$)
\begin{equation}\label{A}
4 \pi \,H(y,y') = \frac{1}{2}\left(\frac{1}{\sigma_+ (\mu - \mu',r,
r',z-z')} - \frac{1}{\sigma_-(\mu - \mu', r, r',z-z')}\right)\;,
\end{equation}
with $\sigma_{\pm}(\mu, r, r',z)$ being implicitly given by
\begin{equation}
\sigma_{\pm} = \pm \,\sqrt{r^2 + r'^2 - 2 r r' \cos(\mu + \Omega
\sigma_{\pm}) + z^2}
\end{equation}
(If we had not chosen the symmetric Green function the quantity corresponding to $H(y,y')$ would have been different,
in particular not symmetric.) The quotient metric $h$ is given by
\begin{equation}
h_{ij}\,dx^i dx^j = d r^2 + \frac{r^2}{1 - \Omega^2 r^2}\; d
\mu^2 + dz^2
\end{equation}
and spacetime Killing vectors projecting to the quotient are given
by $\partial_{x^3} =
\partial_z$ and $\partial_\phi = \partial_\mu$. Applying the main
statement of this Appendix to these two Killing vectors yields the
result concerning the second term in Eq.(\ref{31''}).

\subsection*{Acknowledgement} J\"urgen Ehlers has greatly influenced
our scientific thinking. We dedicate this work to his memory.

\end{document}